\documentstyle[12pt,aps,epsf]{revtex}
\begin{document}
\begin{center}
{\Huge Top-Charm Associated Production at High Energy  $e^{+}e^{-}
$ Colliders in Standard Model}\\
\vspace{1 cm}
{\Large Chao-Shang Huang $^a$, Xiao-Hong Wu $^a$ and Shou-Hua Zhu $^{b,a}$} \\
\vglue 1cm
$^a$ Institute of Theoretical Physics, Academia Sinica, P. O. Box
2735,\\ Beijing 100080, P. R. China \\
$^b$ CCAST (World Lab), P.O. Box 8730, Beijing 100080, P.R. China \\
\vglue 0.5cm
\end{center}
\date{}
\vskip 2.5cm
\begin{center} Abstract \\
\end{center}
\vspace{0.5 cm}
The flavor changing neutral current tcV(V=$\gamma$,Z) couplings in the production vertex 
for the process $e^+e^-\rightarrow t\bar c\mbox{ or }\bar t c$ in the standard model are investigated.
The precise calculations keeping all quark masses non-zero are carried out.
 The total production cross section is found to be  $1.84 \times 10^{-9}$ fb
at $\sqrt s$=200 Gev and  $0.572 \times 10^{-9}$ fb at $\sqrt s$=500 Gev respectively. The result
is  much smaller than that given in 
ref.~\cite{clwy} by a factor of $10^{-5}$.
\newpage

\hspace{4mm}Top quark physics has been extensively investigated~\cite{rev}. The advantage of
examining top quark physics than other quark physics is that one can directly determine
the properties of top quark itself and does not need to worry about non-perturbative QCD
effects which are difficult to attack because there exist no top-flavored hadron states 
at all. The properties of top quark could reveal information on flavor
physics, electroweak symmetry breaking as well new physics beyond the standard
model(SM). 

One of important fields in top physics is to study flavor changing
neutral current (FCNC) coupings. There are no flavor changing
neutral currents at tree-level in the SM. FCNC appear
at loop-levels and consequently offer a good place to test quantum effects of
the fundamental quantum field theory on which SM based. Furthermore, they are
very small at one loop-level due to the unitary of Cabbibo-Kobayashi-Maskawa
 (CKM) matrix. In models beyond SM new particles beyond the particles in SM
may appear in the loop and have significant contributions to flavor
changing transitions. Therefore, FCNC interactions give an ideal place to
search for new physics. Any positive observation of FCNC couplings
deviated from that in SM would unambiguously signal the presence of new
physics. Searching for FCNC is clearly one of important goals of high energy
colliders, in particular, $e^+e^-$ colliders~\cite{pro}.

The flavor changing transitions involving external up-type quarks which are
due to FCNC couplings are much more suppressed than those involving
external down-type quarks in SM. The effects for external up-type quarks
are derived by virtual exchanges of down-type quarks in a loop for which
GIM mechanism~\cite{gim} is much more effective because the mass splittings
between down-type quarks are much less than those between up-type quarks.
Therefore, the tc transition which is studied in the latter
opens a good window to search for new physics.

The FCNC vertices tcV(V=$\gamma$, Z) can be probed either in rare decays
of t quark or via top-charm associated production.
A lot of works have been done in the former case~\cite{rd}. And a number of
papers on the latter case have also appeared~\cite{hh,clwy,many}. In this letter we shall
investigate the latter case in the process \begin{eqnarray}
e^+e^-\rightarrow t\bar{c}\mbox{ or } \bar{t}c. 
\end{eqnarray}
Comparing t quark rare decays where the momentum transfer $q^2$ is
limited, i. e., it should be less or equal to mass square of t quark
$m_t^2$, the production process (1) allows the large (time-like) momentum
transfer, which is actually determined by the energies available at
$e^+e^-$ colliders. The reaction (1) has some advantages because of the
ability to probe higher dimension operators at large momenta and striking
kinematic signatures which are straightforward to detect in the clean
environment of $e^+e^-$ collisions. In particular, in some extensions of
SM which induce FCNC there are large underlying mass scales and large
momentum transfer so that these models are more naturally probed via
t$\bar{c}$ associated production than t quark rare decays. 

The production cross sections of the process (1) in SM have been calculated in
refs.~\cite{clwy,many}. In the early references~\cite{many} a top quark mass
$m_t\le m_Z$ is assumed and the on-shell Z boson dominance is adopted. The
reference~\cite{clwy} considered a large top quark mass and abandoned the 
on-shell Z boson dominance. However, the "self energy" diagrams  have been
omitted in ref.~\cite{clwy}. This is  not legal because the one-loop
contribution for FC transitions is of the leading term of the FC transitions
and must be finite, i.e., although there are some divergences for some
diagrams they should cancel each other in the sum of contributions of all
diagrams. Furthermore, the order of values of cross sections  given in
ref.~\cite{clwy} is not correct.

The order of values of cross sections for the process (1) in SM can easily
be estimated. The differential cross section can be written as
\begin{eqnarray}
\frac{d\sigma}{dcos\theta} ={\frac{N_c}{32\pi s}} (1-\frac{m^2_t}{s})
 {\frac{1}{4}} {\sum_{spins}|M|^2} 
\end{eqnarray}
Where $N_c$ is the color factor, $\theta$ is the the angle between
incoming electron $e^{-}$ and outgoing top quark $t$ and M is the amplitude of
the process. In eq.(2) the charm quark mass in kenetic factors has been
omitted. Due to the GIM mechanism, one has
\begin{eqnarray}
\sum_{spins}|M|^2 & = & e^8 |\sum_{j=d,s,b} V_{jt}^{\star}V_{jc} f(x_j, y_j)|^2
 \nonumber \\ & = &  e^8 |V_{tb}^{\star} V_{cb} \frac{m_b^2-m_s^2}{m_w^2}
\frac {\partial f}{\partial x_j}|_{x_j,y_j=0} + ...|^2,
\end{eqnarray}
where $x_j=m_j^2/m_w^2, y_j=m_j^2/s$, and "..." denote the less important
terms for $\sqrt s\ge 200$ Gev. Assuming $\frac {\partial
f}{\partial x_j}|_{x_j,y_j=0} $ = O(1), one obtains from eqs. (2),(3)\\
$$ \sigma \sim 10^{-8}-10^{-9} fb$$
at $\sqrt s$ = 200 Gev. However, the results given in ref.~\cite{clwy} are\\
$$ \sigma = 0.71\times 10^{-2} fb$$
for $m_t$=165 Gev and
$$ \sigma = 4.1\times 10^{-4} fb$$
for $m_t$=190 Gev, which are much larger than the above estimation by a factor
of $10^{5}$. In order to test SM and search for new physics from observations
of some process one needs to know what are the precise results for the relevant
observables of the process in SM. Therefore, it is necessary to calculate
precisely the cross sections in the SM. In this letter we calculate the differential and
total cross sections  of the process (1) in SM. 
 
In SM for the process (1) there are three kinds of Feynman diagram at one
loop, "self enengy" (actually it is a FC transition, not a usual self energy
diagram), triangle and box diagram, which are shown in Fig.1. We carry out
calculations in the Feynman-t'Hooft gauge. The contributions of the neutral
Higgs H and Goldstone bosons $G^{0,\pm}$ which couple to electrons are
neglected since they are proportional to the electron mass and we have put the
mass of electron to zero.
\par
We do the reduction using FeynCalc ~\cite{3} and keep all masses non-zero except
for the mass of electron. To control
the ultraviolet divergence, the dimensional regularization is used. As a
consistent check, we found that all divergences are canceled in the sum.  The
calculations are carried out in the frame of the centre of mass system (CMS)
and Mandelstam variables have been employed: \begin{eqnarray}
s=(p_1 +p_2)^2 =(k_1 +k_2)^2 \hspace{7mm} t=(p_1 -k_1)^2 \hspace{7mm}
u=(p_1 -k_2)^2,
\end{eqnarray}
 where
$p_1,p_2$ are the
momenta of electron and positron respectively, and $k_1,k_2$ are the momenta of 
top quark $t$ and anti-charm quark $\bar{c}$ respectively.
\par
The amplitude of process $e^{+}e^{-} \rightarrow t\bar{c}$ can be 
expressed as
\begin{eqnarray}
M &=&  \sum_{j=d,s,b} 16\pi^2 \alpha^2 V_{cj}^{\star}V_{tj} [ g_1  \bar{u_t}  \gamma^{\mu}  P_L v_c  \bar{v_e}  \gamma_{\mu}  P_R  u_e  + 
g_2 \bar{u_t}  \gamma^{\mu}  P_L  v_c  \bar{v_e}  \gamma_{\mu}  P_L  u_e  +  g_3 \bar{u_t}  P_L  v_c  \bar{v_e}  \not\!{k_1}  P_R  u_e  +\nonumber\\
&& g_4 \bar{u_t}  P_L  v_c  \bar{v_e}  \not\!{k_1}  P_L  u_e  +  g_5 \bar{u_t} \not\!{p_1}  P_L  v_c \bar{v_e}  \not\!{k_1}  P_L  u_e +
  g_6 \bar{v_e} \gamma^{\mu}  P_L  u_e  \bar{u_t}  \gamma_{\mu}  \not\!{p_1}  P_L  v_c  + \nonumber\\
&& g_7 \bar{u_t} \gamma^{\mu}  P_R  v_c  \bar{v_e}  \gamma_{\mu} P_R  u_e  + 
g_8 \bar{u_t}  \gamma^{\mu}  P_R  v_c  \bar{v_e}  \gamma_{\mu}  P_L  u_e +  g_9 \bar{u_t}  P_R  v_c  \bar{v_e}  \not\!{k_1}  P_R  u_e  +\nonumber\\
&& g_{10} \bar{u_t}P_R  v_c  \bar{v_e}  \not\!{k_1}  P_L  u_e  +  g_{11} \bar{v_e}  \gamma^{\mu}  P_L  u_e  \bar{u_t}  \gamma_{\mu} \not\!{p_1}  P_R  v_c ]
\end{eqnarray}
where $\alpha$ is fine structure constant, $V_{ij}$ is CKM matrix element, $P_L$ is 
defined as $(1-\gamma^5)/2$, and $P_R$ is defined as $(1+\gamma^5)/2$.
The exact expressions of the coefficients $g_j(j=1,2,...11)$ are too long
to be given. Instead, in order to show the essential points, we give them
in the limit of $m_i$/m (i=d,s,c,
m=$m_w, m_t, s)$ approach to zero.  In the limit $g_j(j=7,8,9,10,11)$ is zero,
and the others are given as follows.
\begin{eqnarray}
g_1  &=&  a_3m_j^2 - 2a_4m_j^2s_w^4 + 6a_4C_2^cm_j^2m_t^2s_w^2 + 6m_w^2(2C_{11}^dm_t^2 + 2C_{22}^ds + 2C_{12}^d(m_t^2 + s) )(a_3 + 2a_4c_w^2s_w^2)  +\nonumber\\
&& 12m_w^2C_2^d(a_3(m_t^2 + s) + a_4( 2sc_w^2s_w^2 + m_t^2c_w^2s_w^2 - m_t^2s_w^4)) - B_0^b(m_j^4 - m_j^2m_t^2 + m_j^2m_w^2 + \nonumber\\
&& 2m_t^2m_w^2 - 2m_w^4)(a_1 + 2a_2s_w^2(3 - 4s_w^2)) + 12C_{00}^d(a_3(m_j^2 + 6m_w^2) + a_4s_w^2(c_w^2m_j^2 + 12c_w^2m_w^2 - \nonumber\\
&& m_j^2s_w^2)) + 6C_1^dm_w^2(2a_3(m_t^2 + s) + 2a_4s_w^2(c_w^2m_t^2 + 2c_w^2s - m_t^2s_w^2)) - 6C_0^dm_w^2(2a_3(m_j^2  - s) -\nonumber\\
&& 2a_4s_w^2(2c_w^2m_t^2 + 2c_w^2s + 2m_j^2s_w^2 - m_t^2s_w^2 -3m_t^2c_w^2 )) + 2C_0^cm_j^2(a_3(m_j^2 + 2m_w^2 - m_t^2) + \nonumber\\
&& a_4s_w^2(3m_j^2 - 2m_j^2s_w^2 - 4m_w^2s_w^2 + 2m_t^2s_w^2)) - 2(2C_{00}^c + C_{11}^cm_t^2 + C_{22}^cs + C_2^cs + \nonumber\\
&& C_{12}^c(m_t^2 + s))(a_3(m_j^2 + 2m_w^2) + 2a_4s_w^2(3m_w^2 - m_j^2s_w^2 - 2m_w^2s_w^2)) + \nonumber\\
&& B_0^a( a_1(m_j^2 - m_w^2)(m_j^2 + 2m_w^2) -2a_1m_t^2m_j^2 + 2a_2s_w^2(3m_j^4 - 6m_j^2m_t^2 + 3m_j^2m_w^2 - 6m_w^4 -\nonumber\\
&&  4m_j^4s_w^2 - 4m_j^2m_w^2s_w^2 + 8m_w^4s_w^2 + 8m_t^2m_j^2s_w^2 )) - 2C_1^c(a_3( 2m_w^2s +m_t^2m_j^2) - \nonumber\\
&& a_4s_w^2(3m_j^2m_t^2  + 2m_t^2m_j^2s_w^2 - 6m_w^2s  + 4m_w^2ss_w^2))  
\end{eqnarray}
\begin{eqnarray}
g_2  &=&   a_3m_j^2 + a_4m_j^2(1 - 2s_w^2)(s_w^2 - 3C_2^cm_t^2) + a_5(8D_{00}^e +u(D_1^e +D_2^e +2D_3^e +2D_{12}^e +4D_{13}^e +\nonumber\\
&& 2D_{23}^e +2D_{33}^e) + (2m_t^2D_3^e +2sD_{13}^e +2D_{33}^em_t^2) ) + 6m_w^2(2C_{11}^dm_t^2 + 2C_{22}^ds +\nonumber\\
&& 2C_{12}^d(m_t^2 + s))(a_3 - a_4c_w^2(1 - 2s_w^2))  + 6m_w^2C_2^d(2a_3(m_t^2 +s) - a_4(1-2s_w^2)(m_t^2c_w^2 -m_t^2s_w^2 +\nonumber\\
&& 2sc_w^2)) - B_0^b(m_j^4 - m_j^2m_t^2 + m_j^2m_w^2 + 2m_t^2m_w^2 - 2m_w^4)(a_1 - a_2(3 - 4s_w^2)(1 - 2s_w^2)) + \nonumber\\
&& 6C_{00}^d(2a_3(m_j^2 + 6m_w^2) - a_4(1 - 2s_w^2)(c_w^2m_j^2 + 12c_w^2m_w^2 - m_j^2s_w^2)) + 6C_1^dm_w^2(2a_3(m_t^2 + s) - \nonumber\\
&& a_4(1 - 2s_w^2)(c_w^2m_t^2 + 2c_w^2s - m_t^2s_w^2)) - 6C_0^dm_w^2(2a_3(m_j^2  - s) + a_4(1 - 2s_w^2)(- c_w^2m_t^2 + 2c_w^2s + \nonumber\\
&& 2m_j^2s_w^2 - m_t^2s_w^2)) + C_0^cm_j^2(2a_3(m_j^2 + 2m_w^2 - m_t^2) - a_4(1 - 2s_w^2)(3m_j^2 - 2m_j^2s_w^2 - 4m_w^2s_w^2 +\nonumber\\
&& 2m_t^2s_w^2)) - 2(2C_{00}^c + C_{11}^cm_t^2 + C_{22}^cs + C_2^cs + C_{12}^c(m_t^2 + s))(a_3(m_j^2 + 2m_w^2) - \nonumber\\
&& a_4(1 - 2s_w^2)(3m_w^2 - m_j^2s_w^2 - 2m_w^2s_w^2)) + B_0^a(a_1(m_j^2 - m_w^2)(m_j^2 + 2m_w^2) - 2a_1m_t^2m_j^2 - \nonumber\\
&& a_2(1 - 2s_w^2)(3m_j^4 - 6m_j^2m_t^2 + 3m_j^2m_w^2 - 6m_w^4 - 4m_j^4s_w^2 - 4m_j^2m_w^2s_w^2 + 8m_w^4s_w^2 + 8m_t^2m_j^2s_w^2)) - \nonumber\\
&& C_1^c(2a_3(m_t^2m_j^2 + 2sm_w^2) + a_4(1 - 2s_w^2)(3m_j^2m_t^2  - 6sm_w^2  + 4sm_w^2s_w^2 + 2m_t^2m_j^2s_w^2))  
\end{eqnarray}
\begin{eqnarray}
g_3  &=&  12a_4s_w^2m_t(2C_2^dm_w^2 -C_2^cm_j^2) + 4m_tC_0^cm_j^2(a_3 - 2a_4s_w^4) + 24m_tm_w^2(2C_1^d + C_0^d)(a_3 + 2a_4c_w^2s_w^2)  +\nonumber\\
&& 8C_1^cm_t(a_3m_j^2 - 2a_4s_w^4m_j^2) + 12m_t(C_{11}^d +C_{12}^d)(a_3(m_j^2 + 2m_w^2) + a_4s_w^2(c_w^2m_j^2 + 4c_w^2m_w^2 - m_j^2s_w^2)) +\nonumber\\
&& 4m_t(C_{11}^c + C_{12}^c)(a_3(m_j^2 + 2m_w^2) + 2a_4s_w^2(3m_w^2 - m_j^2s_w^2 - 2m_w^2s_w^2))   
\end{eqnarray}
\begin{eqnarray}
g_4  &=&  - 2a_5m_t(2D_{23}^e + D_2^e + 2D_{33}^e + 2D_3^e) + 6a_4m_t(C_2^cm_j^2 -2C_2^dm_w^2)(1 - 2s_w^2) +  4C_0^cm_tm_j^2(a_3 + \nonumber\\
&& a_4s_w^2(1 - 2s_w^2)) + 24m_tm_w^2(2C_1^d + C_0^d )(a_3 - a_4c_w^2(1 - 2s_w^2))  +  8C_1^cm_t(a_3m_j^2 + \nonumber\\
&& a_4s_w^2m_j^2(1 - 2s_w^2)) + 6(C_{11}^d + C_{12}^d)m_t(2a_3(m_j^2 + 2m_w^2) - a_4(1 - 2s_w^2)(c_w^2m_j^2 + 4c_w^2m_w^2 - \nonumber\\
&& m_j^2s_w^2)) + 4m_t(C_{11}^c +C_{12}^c)(a_3(m_j^2 + 2m_w^2) - a_4(1 - 2s_w^2)(3m_w^2 - m_j^2s_w^2 - 2m_w^2s_w^2))    \\
g_5  &=&  - 4a_5( D_{12}^e + D_{13}^e )  \\
g_6  &=& a_5m_t( 2D_{12}^e + 2D_{13}^e  + 2D_{23}^e + D_2^e +2D_{33}^e + 2D_3^e)
\end{eqnarray}
with $m_j^2=m_b^2$(since $m_s,m_d$ have been omitted in the above
expressions of g's),\\
where $a_i (i=1,2,...,5)$ are defined by
\begin{eqnarray}
a_1=\frac{1}{96s\pi^2s_w^2m_t^2m_w^2},\hspace{3mm} a_2=\frac{1}{768\pi^2c_w^2s_w^4m_t^2m_w^2 (m_z^2-im_z\Gamma{z}-s)},
 \hspace{3mm} a_3=\frac{1}{192s\pi^2s_w^2 m_w^2}  \nonumber\\
 a_4=\frac{1}{384\pi^2c_w^2s_w^4m_w^2 (m_z^2-im_z\Gamma{z}-s)},\hspace{3mm} a_5=\frac{1}{32\pi^2s_w^4} \nonumber
\end{eqnarray}
with $ c_w= cos\theta_w$ and $s_w= sin\theta_w$.
In the presentation of $g_j$ above, we have used the definition of scalar
integrals $Bs$, $Cs$,and $Ds$\cite{3}, and these
functions, $Bs$, $Cs$,and $Ds$, with superscripts a,b,...,e have the arguments
\\ 
\begin{math}
(0,m_j^2,m_w^2),\hspace{3mm} (m_t^2,m_j^2,m_w^2),\hspace{3mm} 
(m_t^2,0,s,m_j^2,m_w^2,m_j^2), \hspace{3mm} (m_t^2,0,s,m_w^2,m_j^2,m_w^2)\\
\hspace{3cm} (0,s,m_t^2,u,0,0,0,m_w^2,m_w^2,m_j^2)\\
\end{math}
respectively. Here $m_j$ denotes the mass of down-type quark b.

\hspace{4mm} In the numerical calculations the following values of the
parameters have been used ~\cite{4}: 
\begin{eqnarray}
m_e=0,\hspace{4mm}m_c=1.4Gev,\hspace{4mm}m_t=175Gev, \hspace{4mm}m_d=0.005Gev,\hspace{4mm}m_s=0.17Gev,\hspace{4mm}\nonumber\\
m_b=4.4Gev,\hspace{4mm}m_w=80.41Gev,\hspace{4mm}m_z=91.187Gev,\hspace{4mm} \Gamma_z=2.5Gev,\hspace{4mm} \alpha=\frac{1}{128} \nonumber
\end{eqnarray}
\par
In order to keep the unitary condition of CKM matrix exactly, we employ the 
standard parametrization and take the values~\cite{4,ckm}
\begin{eqnarray}
s_{12}=0.220,\hspace{7mm} s_{23}=0.039,\hspace{7mm}
s_{13}=0.0031,\hspace{7mm} \delta_{13}=70^{\circ} . \nonumber
\end{eqnarray}

Numerical results are shown in Figs. 2, 3.
In Fig.2, we show the total cross section $\sigma_{tot}$ of the process
$e^{+}e^{-} \rightarrow t\bar{c}$ as a function of the centre of mass energy
$\sqrt{s}$.  One can see from the figure that the
total cross section is the order of $10^{-10} \sim 10^{-9}$ fb, as expected, and decreases when center-of-mass energy
increases and is large enough ($\ge 250$ Gev ).
We fixed the centre of mass energy $\sqrt{s}$ at $200 Gev$. Differential cross 
section of the process at the energy as a function of $\cos{\theta}$
 is shown in Fig.3.
\par 
To summarize, we have calculated the production cross sections of the process
$e^{+}e^{-} \rightarrow t\bar{c}$ in SM.  We found that the total cross
section is $1.84\times 10^{-9} fb$ at $\sqrt s$ = 200 Gev and $0.572\times 10^{-9}$ fb at
$\sqrt s$ = 500 Gev. It is too small to be of experimental relevance.
Therefore, this is a remarkable situation that allows for a precise test of the
SM and, in particular, of the GIM mechanism in SM. Even a small number of
$t\bar c$ events, detected at LEP II or a NLC running with a yearly integrated
luminosity of ${\cal L}\ge 10^2 [fb]^{-1}$, will unambiguously indicate new
FCNC dynamics beyond SM. \\

\section*{\Large Acknowledgements}

This research was supported in part by the National Nature Science
Foundation of China and the post doctoral foundation
of China. One of authors (C. S. Huang) would like to thank IISc and JNCASR
where part of the letter was written for warm hospitality and ICTP for the ICTP-IISc-JNCASR
Associateship Programme.
S.H. Zhu gratefully acknowledges the
support of  K.C. Wong Education Foundation, Hong Kong.


\begin{figure}[p]
\epsfxsize=13cm
\centerline{\epsffile{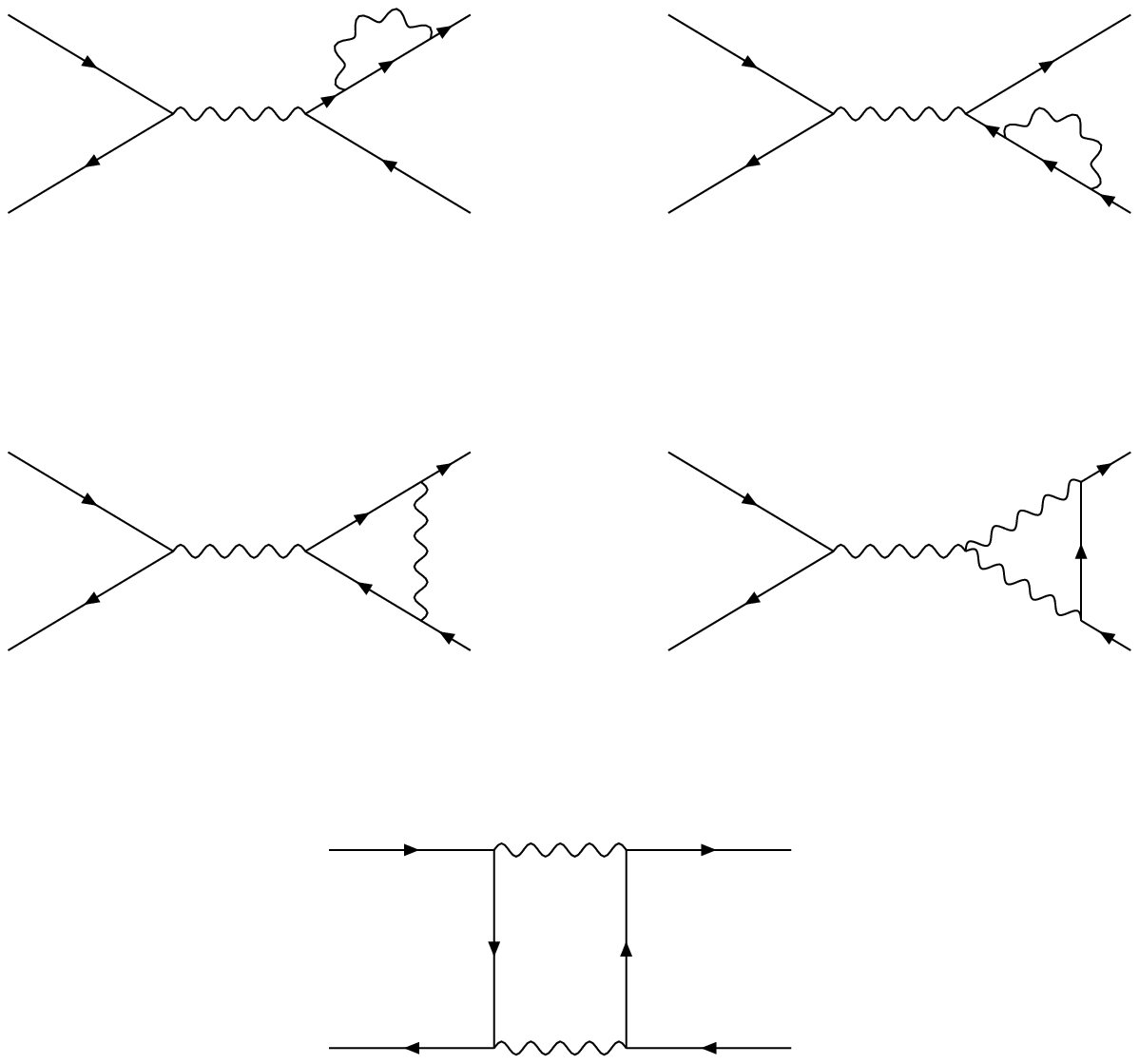}}
\caption[]{Feynman diagrams of
prosess $e^+e^- \rightarrow t \bar c$} 
\end{figure}

\begin{figure}
\epsfxsize=15cm
\centerline{\epsffile{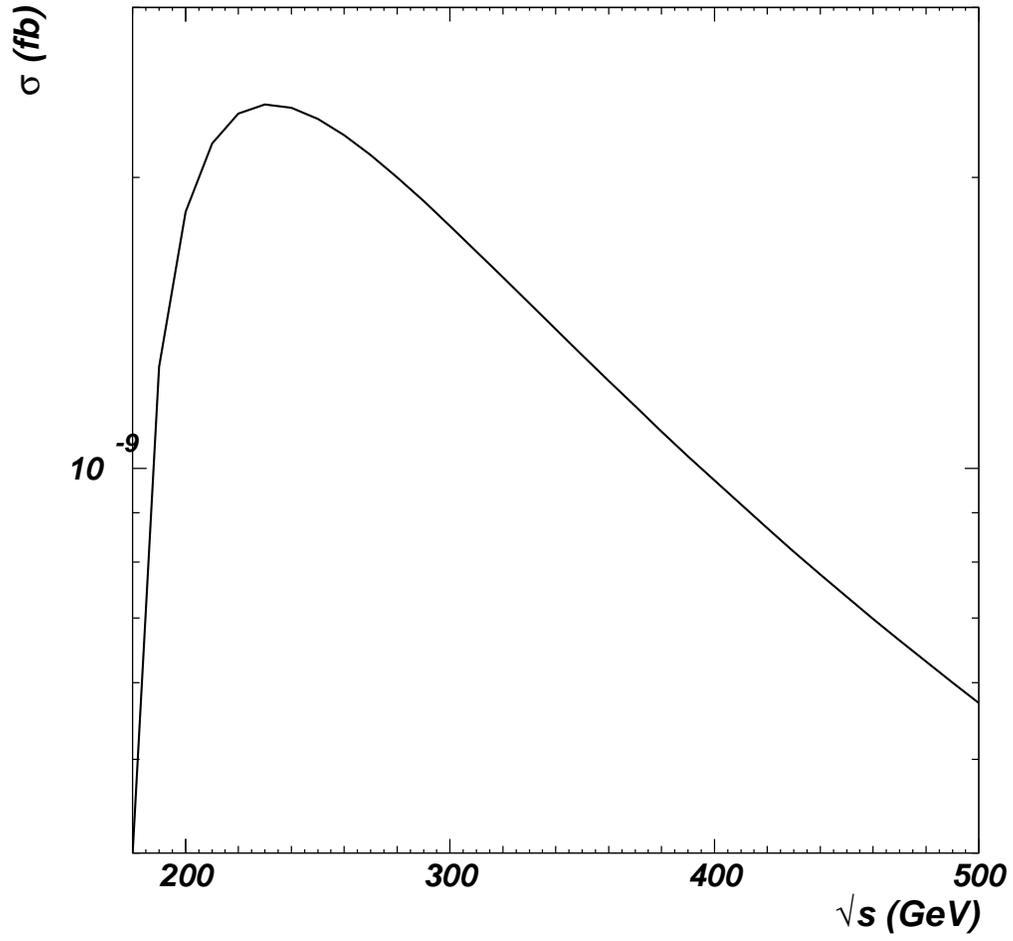}}
\caption[]{ Cross section of the 
process $e^+e^- \rightarrow t \bar c$ as a function of
$\sqrt{s}$.}
\end{figure}

\begin{figure}
\epsfxsize=18cm
\centerline{\epsffile{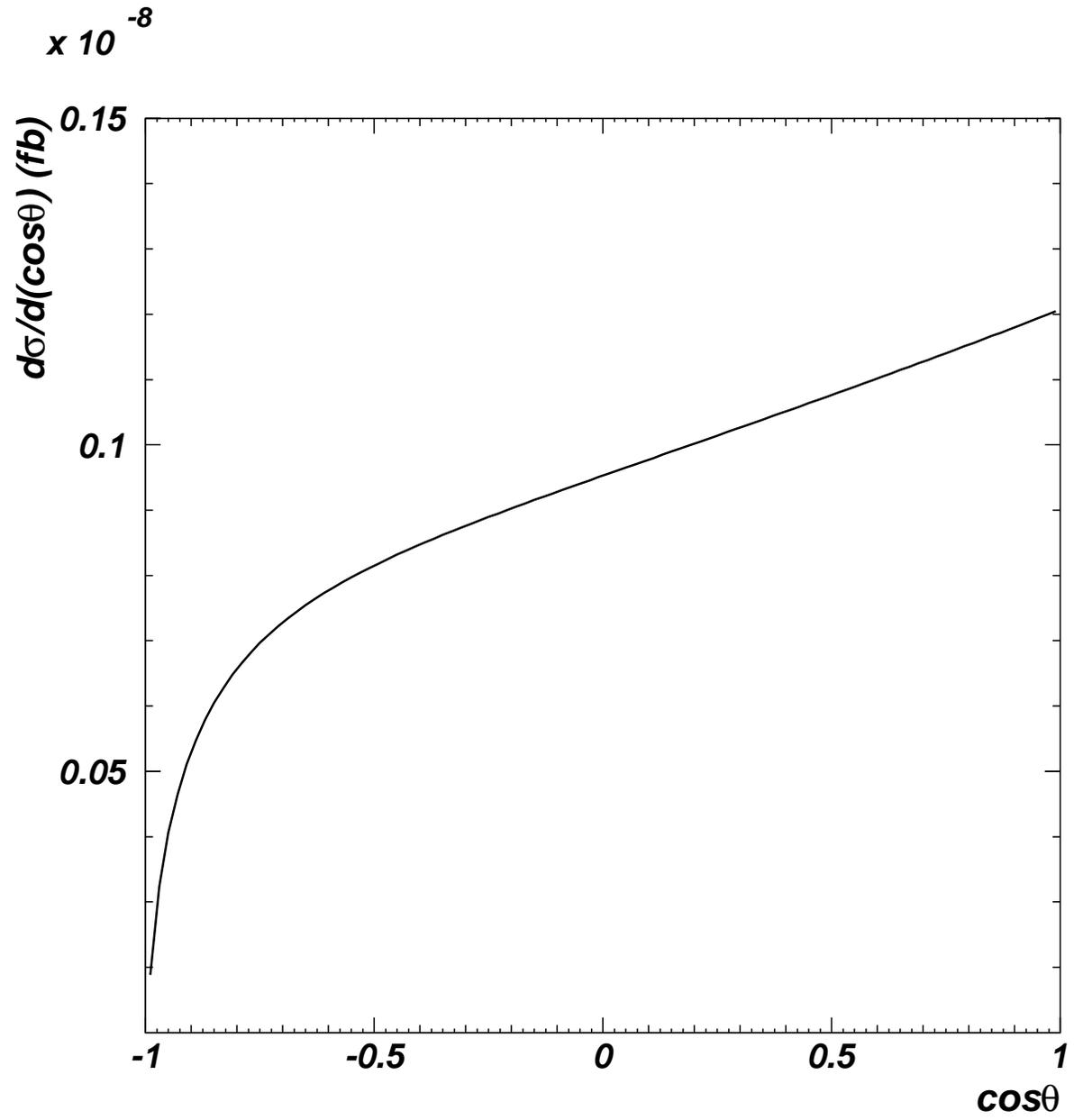}}
\caption[]{  Differential cross section of the 
process $e^+e^- \rightarrow t \bar c$, where
$\sqrt{s}=200$ GeV. 
}
\end{figure}

\end{document}